\documentclass[pdflatex,sn-mathphys-num]{sn-jnl}

\usepackage{graphicx}%
\usepackage{multirow}%
\usepackage{amsmath,amssymb,amsfonts}%
\usepackage{amsthm}%
\usepackage{mathrsfs}%
\usepackage[title]{appendix}%
\usepackage{xcolor}%
\usepackage{textcomp}%
\usepackage{manyfoot}%
\usepackage{booktabs}%
\usepackage{algorithm}%
\usepackage{algorithmicx}%
\usepackage{algpseudocode}%
\usepackage{listings}%
\usepackage{tikz}
\usepackage{pgfplots}
\usepackage{circuitikz}
\usetikzlibrary{patterns}

\theoremstyle{thmstyleone}%
\theoremstyle{thmstyletwo}%

\theoremstyle{thmstylethree}%

\begin{document}

\title[Article Title]{Accuracy in the Measurement of Power Balance in Bicycle Power Meters}


\author*[1]{\fnm{Jack} \sur{Renshaw}}\email{j.renshaw@unsw.edu.au}

\affil*[1]{\orgname{BEng (Hons), The University of New South Wales}, \country{Australia}}

\abstract{Bicycle Power Meters have become ubiquitous as training aids in recent years. Many power meters purport to measure Left-Right Pedal Balance, which is a measure of the extent to which power generated by the application of torque to the left and right pedal differ. This metric has both practical and clinical significance. Most Bicycle Power Meters measure only the Average Angular Velocity (AAV) of the pedal throughout a given pedal stroke - only a small number of power meters compute power using Instantaneous Angular Velocity (IAV). The paper highlights that power balance figures reported by devices employing these two methods have different physical meanings, and that AAV-based power balance figures will tend to be inaccurate when phase imbalance exists within a pedal stroke. Simulations were performed on a number of realistic indoor and outdoor riding scenarios. This results of this paper indicate that errors in Power Balance of up to 10\% could occur under certain conditions.}


\maketitle

\section{Introduction}\label{sec1}
Modern cyclists utilise Bicycle Power Meters as training aids, to measure their efforts in bicycle racing, and to track changes in fitness. Given the minuscule margins that can make or break victory in the sport, the accuracy of these devices is of the upmost importance. 
Some Power Meters measure the net torque produced by the application of
torque to the pedals, whereas other Power Meters measure torque on both the left and right hand side of the bicycle, either at the crank arm or at the pedal \cite{s22010386}.
When torque is measured on both pedals, these independent \textit{Unilateral Torque} measurements may be combined to form a metric called "L/R Power Balance" (Balance).
Balance is not a measure of performance, however the asymmetrical application of power 
could be indicative of an injury \cite{bib14} \cite{rannama2016pedalling}. It has been hypothesised that 
asymmetrical pedalling could also result in injury, however there
is not sufficient evidence in the extant literature to draw a causal connection 
between asymmetrical pedalling and injury\cite{rannama2016pedalling}. \\
\\
Given the growing interest in Balance, this paper will seek to clarify the precise 
meaning of the term given the disparate designs of Power Meters currently available.
This paper will highlight that the physical meaning of Balance differs between 
Power Meters designs, and this difference may have clinical significance - 
in particular, devices that compute power by assuming an constant Angular Velocity, which is termed the \textit{Average Angular Velocity} (AAV), as opposed to \textit{Instantaneous Angular Velocity} (IAV) may produce differing balance figures.\\
\\
The \textit{Favero Assioma} computes power using IAV\cite{faveroproduct}, whereas the Garmin Vector\cite{garminproduct} and Shimano Power Meters\cite{shimanoproduct}
compute power using AAV. The Favero Assioma Power Meters
have been assessed\cite{ValidityoftheFaveroAssiomaDuoPowerPedalSysteminMaximalEffortCyclingTests} as a valid tool for measuring cycling dynamics, and the Garmin Vector has been used in previous research to investigate
the impact of asymmetry on power balance and injury \cite{rannama2016pedalling}. A recent systematic scoping review has noted issues surrounding
the commensurability of power meter readings, however the impact of IAV and AAV was not noted in terms of the impact on power balance\cite{s22010386}.\\
\\
To this end, this paper will first define power, phase symmetry, and Balance
in the context of cycling in terms of the torque
applied by a cyclist at the pedal, and the angular
velocity of the crankarm about its axis.
An idealised model of the pedal stroke of a typical cyclist will be introduced
in terms of applied torque, and the relationship between
applied torque and angular velocity will be described under conditions of \textit{Dynamic Equilibrium}. 
An analytical relationship between phase symmetry and parameters
that the \textit{Bicycle-Rider System} (BRS) and the external environment
through which the BRS transits, will be derived.
Finally, this paper will demonstrate
that reasonable deviations from phase symmetry will result in inaccurate
Balance readings.
\subsection{Cycling Power}
Power, broadly defined, is the rate at which mechanical work is done by the cyclist.
Cyclists perform work against wind resistance, friction and gravity, 
and work is performed by applying a \textit{torque}, $\tau$, to the pedal,
which is rotating about the bicycle bottom bracket at a certain rate, $\omega$. Mechanical linkages
translate this torque into a force applied by the \textit{Bicycle-Rider System} (BRS) against external forces.
\begin{equation}
  P = \tau \times \omega \approx F * v
\end{equation}
The average power over the period of a single pedal stroke is:
\begin{equation}
  P_{avg} = \frac{1}{T} \int_{0}^{2 \pi} \tau(\theta) \times \omega(\theta) \: d \theta = \frac{1}{T} \int_{0}^{T} \tau(t) \times \omega(t) \: dt
\end{equation}
For Power Meters employing AAV compute power by sampling left and right torque $\tau_{l}$ \& $\tau_{r}$,
time-difference is used compute average angular velocity ($\omega$)\cite{s22010386}.
An approximation (approximations are denoted herein by subscript $\sim$) for power is given by:
\begin{equation}
  P_{avg,\sim} = P_{l,\sim} + P_{r,\sim} = \sum_{t=0}^{N} \frac{\tau_{l}+\tau_{r}}{N*(T_{0}-T_{-1})}
\end{equation}
This method yields an approximation of actual power, due to the fact that 
this method neglects power produced by harmonics of torque and angular velocity\cite{faverostudy}.
Whether or not this approximation yields meaningful total error ($P_{avg,\sim} \approx P_{avg}$)
is not the subject of this paper\footnote{In general, the fact that inertial impedance (i.e. reactance) is high tends to result in the harmonic components of net torque and angular velocity being in quadrature, and thus not interacting to produce any real work},
but has been previously investigated\cite{faverostudy}\cite{renshaw2024truly}.
The existence of a \textit{Phase Offset} in \textit{Unilateral Torque} may result
in \textit{Unilateral Power Error} - defined as the difference between the
power produced by the application of torque to one pedal and the erroneous approximation - and thus Balance Error.
The adopted working definition of power balance is normalised 
difference between left and right pedal power.
\begin{equation}
  Unbalance = \frac{P_{l}-P_{r}}{P_{l}+P_{r}} \approx Unbalance_{\sim} = \frac{P_{l,\sim} - P_{r,\sim}}{P_{l,\sim}+P_{r,\sim}}
\end{equation}
The primary question of this paper is to determine 
under what conditions, if any, $Unbalance$ and $Unbalance_{\sim}$
meaninfully diverge. There are two drivers of divergence:
\begin{enumerate}
  \item The pedal stroke of a cyclist, characterised in terms of torque application in Section \ref{subsec:pedalmodel}; and
  \item The interaction of the pedal stroke with the dynamics of the BRS, outlined in Section \ref{sec:dynamicmodel}.
\end{enumerate}
\subsection{Pedal Stroke Model}\label{subsec:pedalmodel}
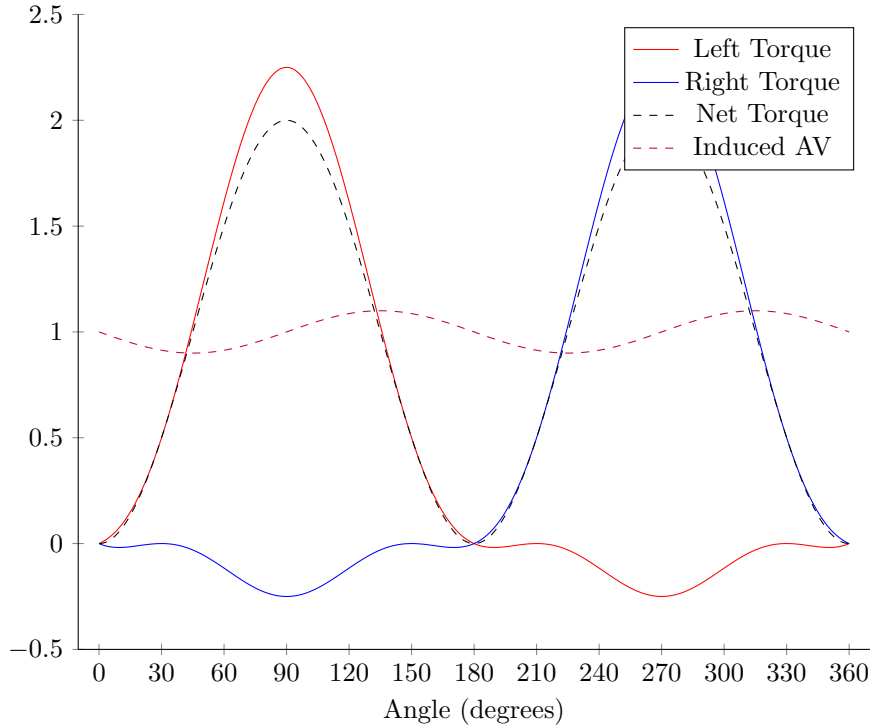
\begin{figure}[H]
    \centering
    \begin{tikzpicture}
      \begin{axis}[
        axis lines=left, 
        axis line style={-}, 
        xmin=-10, xmax=370, 
        ymin=-0.5, ymax=2.5, 
        width=12cm, height=10cm, 
        axis on top, 
        clip=false, 
        xlabel={Angle (degrees)}, 
        xtick={0,30,...,360},
      ]
          
      \pgfmathsetmacro\Aone{75}
      \pgfmathsetmacro\sigmaone{25}
      \pgfmathsetmacro\inertia{5}
      \pgfmathsetmacro\basicv{20}
      \pgfmathsetmacro\offset{10}
  
      \pgfmathsetmacro\Atwo{5}
      \pgfmathsetmacro\sigmatwo{25}

      \addplot[domain=0:360, samples=1000, red, smooth] {0.5+1.0*cos(x-90)+0.5*cos(2*x+180)+0.25*cos(3*x+90)};
      \addlegendentry{Left Torque};
      \addplot[domain=0:360, samples=1000, blue, smooth] {0.5+1.0*cos(x+90)+0.5*cos(2*x+180)+0.25*cos(3*x-90)};
      \addlegendentry{Right Torque};
      \addplot[domain=0:360, samples=1000, black, dashed] {1.0+1.0*cos(2*x+180)};
      \addlegendentry{Net Torque};
      \addplot[domain=0:360, samples=1000, purple, dashed] {1.0+0.1*cos(2*x+90)};
      \addlegendentry{Induced AV};
    \end{axis}
    \end{tikzpicture}
    \caption{Variations in Applied Torque}\label{fig:pedalstroke}
  \end{figure}

There is a large degree of variance in the pedal stroke - defined by the
periodic application of torque - between cyclists.
In the interests of generality, this paper shall characterise the pedal stroke of a typical cyclist
in terms of the minimum number of constituent frequency components
that can adequately describe the basic characteristics of the typical pedal stroke\cite{REDFIELD1986523}.
This paper therefore assumes periodicity with respect to the application of torque.
These essential characteristics are:
\begin{enumerate}
  \item For a Symmetrical Cyclist, Unilateral Torque exhibits a single peak at roughly $90^{\circ}$ degrees and $-90^{\circ}$ degrees for Left and Right Torque respectively.
  \item There is a corresponding negative minimum of Unilateral torque $180^{\circ}$ following peak torque.
  \item Net Torque Exhibits two primary peaks of comparable magnitude, and two \textit{Dead Spots} of zero torque at $0^{\circ}$ degrees and $180^{\circ}$ degrees.
\end{enumerate}
Phase Asymmetry is investigated via the introduction of a \textit{Phase Offset} for
Right Torque from its symmetrical point at $-90^{\circ}$ - labelled ($\alpha$).
Left and Right Torque are constructed to have equal magnitude. 
This means that $Unbalance_{\sim} = 0.0$, and thus $\epsilon_{balance} = Unbalance$.
Mathematically, the expressions for Left and Right Torque are defined in Equations \ref{eq:lefttorque}
and \ref{eq:righttorque}.
\begin{equation}
  \tau_{l}(t) = \frac{1}{2} + \cos(2\pi ft + \frac{\pi}{2}) + \frac{\cos(4\pi ft + \pi)}{2} + \frac{\cos(6\pi ft + \frac{\pi}{2})}{4}
  \label{eq:lefttorque}
\end{equation}
\begin{equation}
  \tau_{r}(t) = \frac{1}{2} + \cos(2\pi ft - \frac{\pi}{2}+\alpha) + \frac{\cos(4\pi ft + \pi)}{2} + \frac{\cos(6\pi ft - \frac{\pi}{2})}{4}\\
  \label{eq:righttorque}
\end{equation}
\\
Noting that $\cos(x + a) + \cos(x + b) = 2\cos(x+\frac{a+b}{2})\cos(\frac{a-b}{2})$, 
and $\cos(\frac{\pi}{2}+x) = \sin(x)$,
Net Torque, $\tau_{t}(t) = \tau_{l}(t) + \tau_{r}(t)$, is thus:
\begin{equation}
  \tau_{t}(t) = 1 + \cos(2\pi ft + \frac{\alpha}{2})\sin(\frac{\alpha}{2}) + \cos(4\pi ft + \pi)
  \label{eq:nettorque}
\end{equation}
Under steady conditions, \textit{Net Torque}
will a certain Angular Velocity (AV). Variations 
in Net Torque will thus manifest as variations in AV, with some degree
of damping (due to cyclist inertia provided by mass), and a phase offset, $\delta_{n}$, with respect
to torque for each frequency component (i.e. the $n_{th}$ harmonic of the fundamental).\\
\\
This is visualised in Figure \ref{fig:pedalstroke}
for the special case where $\alpha = 0$, which results
in the destructive interference of the first and third harmonics of
torque. In this instance, there is only a single frequency component 
of net torque and AV.\\
Observe that AV is periodic. This is not true in general\footnote{AV will not be periodic if there is net acceleration during the pedal stroke, either due to high force or changing external forces},
however AV is treated as periodic for the remainder of this paper due to the assumption
of Dynamic Equilibrium section in Sub-Section \ref{sub:dynamicequil}. Also 
note that AV lags torque by $\approx 90^{\circ}$.
The precise relationship between Torque and AV is characterised in Section \ref{sec:dynamicmodel}.
  \section{Dynamic Model of a Cyclist}\label{sec:dynamicmodel}
The section will outline the model used for characterising the behaviour of a \textit{Bicycle-Rider System} (BRS)
when the system is subject to varying Left and Right Pedal Torques. 
This Dynamic Model is based on the model previously developed by Renshaw\cite{renshaw2024truly}
to assess the impact of \textit{IAV} and \textit{AAV} on Power Meter Accuracy.
\subsection{A Mechanical and Kinematic Model of a Bicycle-Rider System}
A cyclist travelling at linear velocity $v$ and up a gradient with angle $\theta$ is subject to wind resistance, a quadratic function of the BRS velocity\footnote{Assuming there is no headwind},
a force due to gravity, and linear friction (from tyre rolling resistance, for example)\cite{cangley2010modelling}.
This can be expressed in Equation \ref{eq:kinematicforces}.
\begin{equation}
  F_{t} = \frac{1}{2} \rho CdAv^2 + k_{f}v + mgsin(\theta)
  \label{eq:kinematicforces}
\end{equation}
Where:
\begin{description}
  \item[$v$] Is the linear velocity of the bicycle-rider system with respect to the ground in the direction of travel.
  \item[$m$] Is the mass of the entire bicycle-rider system.
  \item[$g$] Is the gravitational constant.
  \item[$\rho$] Is air density.
  \item[$Cd$] Is the coefficient of drag of the entire system.
  \item[$A$] Is the frontal surface area of the entire system.
  \item[$k_{f}$] Is the coefficient of linear velocity-dependent friction(s). 
  \end{description}
For the purposes of this paper, a linear approximation is taken for all resistive 
forces, labelled \textit{friction}, which gives rise to the following equation of motion:
\begin{equation}
  F = mg \sin(\theta) + k \dot{x} + m \ddot{x}
  \label{eq:motion}
\end{equation}
Wind Resistance, quadratically varying with velocity,
tends to be the dominant frictional term at higher velocities. For small values
around the average linear velocity of the cyclist, $v_{avg}$, the linearised coefficient of wind 
resistance is:
\begin{equation}
  k = \frac{1}{2} \rho CdAv_{avg}
  \label{eq:windcoeff}
\end{equation}
\subsection{A Cyclist in Dynamic Equilibrium}\label{sub:dynamicequil}
A cyclists velocity on a bicycle ride, $v(t)$,
will contain some amount of variation.
The angular velocity 
of the pedal, $\omega(t)$, will also vary and,
in a constant gear, $v(t) \propto \omega(t)$\footnote{This is assuming that bicycle gears are perfectly stiff and circular, that force and velocity are seamlessly transferred from the pedal to the rear wheel.}.
This paper distinguishes between two types of variation in velocity:
\begin{enumerate}
  \item \textbf{Oscillations}, which are periodic variations in velocity that are contained within a single pedal stroke. As shown in Figure \ref{fig:pedalstroke}, these Oscillations are \textit{induced} by Oscillations in Torque/Force, and thus \textit{lag} Torque in Angular Position and Time.
  \item \textbf{Accelerations}, which are net changes in velocity from the start to the end of a pedal stroke.
\end{enumerate}
If net acceleration over some time period $T$ (the length of a pedal stroke) is zero, then a cyclist is in \textit{Dynamic Equilibrium} over that period.
The scope of this paper is limited only to scenarios of dynamic equilibrium.
Practically, this replicates most conditions under which Power Balance has
clinical significance or is measured\cite{rannama2016pedalling}.
  \subsection{System Modelling}
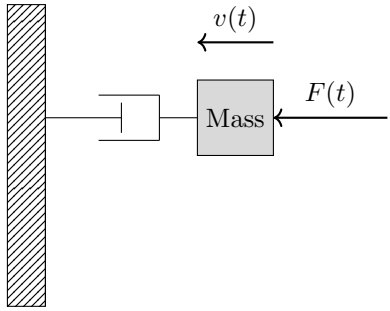
\begin{figure}[H]
    \centering
    \begin{tikzpicture}
    \draw (0,2) -- (0,-2);
    \draw[pattern=north east lines, pattern color=black] (-0.5,2) rectangle (0,-2);
    \draw[fill=gray!30] (2,0) rectangle node[text width=2cm,align=center] {Mass} (3,1);
      
      \draw[->,thick] (4.5,0.5) -- node[above]{$F(t)$} (3,0.5);
      \draw[->,thick] (3,1.5) -- node[above]{$v(t)$} (2,1.5);
    \draw (0,0.5) -- (1,0.5);
    \draw (1,0.3) -- (1,0.7);
    \draw (0.7,0.2) -- (1.5,0.2);
    \draw (0.7,0.8) -- (1.5,0.8);
    \draw (1.5,0.2) -- (1.5,0.8);
    \draw (1.5,0.5) -- (2.0,0.5);

    \end{tikzpicture}
    \caption{Mass-Damper Model of a Cyclist}\label{fig:massdamper}
  \end{figure}

\begin{figure}[H]
    \centering
      \begin{circuitikz}
      \draw (0,0) to[V, v=$V_{l}$] (0,2)
                  to[V, v=$V_{r}$] (0,4)
                  to[V, v<=$V_{g}$, invert] (0,6)
                  to[L, l=$L$] (4,6)
                  to[R, l=$R$] (4,0)
                  -- (0,0);
                  \draw (-0.3,4.0) node[label={[font=\footnotesize, below]:\textbf{$V_{t}$}}]{};
                  \draw (0,6) -- ++(0.5,0) coordinate (currentarrowstart);
                  \draw[-{Latex[width=2mm]}] (currentarrowstart) -- ++(0.5,0) node[above, right] {\textbf{$I_{t}$}};
  \end{circuitikz}
\caption{General Circuit Model of a BRS}\label{fig:generalcircuit}
\label{fig:cyclistcircuit}
\end{figure}
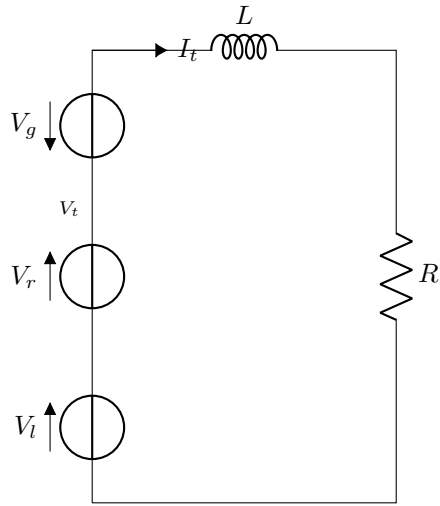

  The BRS can be modelled as a mass-damper system, shown in Figure \ref{fig:massdamper},
  based on the Equation of Motion set of out in Equation \ref{eq:motion}.
The sources of impedance (to movement)
can be replaced with their electrical analogues (Inductance
for Mass, and Resistance for Friction and Wind Resistance). 
Employing that same analogy, the Force applied by the system
(which is transferred from pedal force via gearing) is represented
by Voltage, and the resultant\footnote{Resultant, because under the system as modelled Current (or Velocity/Angular Velocity), is entirely a consequence of Voltage (or Force/Torque)} 
Current represents Velocity.
This is shown in Figure \ref{fig:generalcircuit} - 
$V_{l}$ and $V_{r}$ represent the equivelant system force
resulting from Left and Right pedal Torque Respectively, assuming
a constant gear ratio. $V_{t}$
represents the net applied force, the summation
of $V_{l}$ and $V_{r}$, and $V_{g}$ represents
the constant opposing force due to gravity. $I_{t}$
represents the velocity of the system.
The Force Due to Gravity is given by:
\begin{equation}
  V_{g} = mg\sin(\theta)
\end{equation}
The value of resistance, adopted from Equation \ref{eq:windcoeff}, is as follows
\begin{equation}\label{eq:resistance}
  R = \frac{1}{2} \rho CdAv_{avg}
\end{equation}
Reactive Impedance resulting from mass is frequency dependant.
For the $i_{th}$ harmonic, the Reactive Impedance is given by:
\begin{equation}
  X_{i} = j \omega m
\end{equation}
The general form for $V_{l},V_{r}$ and $V_{t}$ is:
\begin{equation}
  V(t) = V_{0} + \sum_{i=1}^{3} V_{n} \cos(2i\pi ft + \phi_{n})
\end{equation}
At steady state\footnote{i.e. neglecting transient effects}, the resultant current flowing in 
the circuit takes the following form due to the Principle of Superposition\footnote{Essentially, the $i_{th}$ voltage harmonic alone will induce an $i_{th}$ harmonic current. The magnitude and phase of this $i_{th}$ harmonic current depends only on the magnitude of the $i_{th}$ harmonic voltage and the impedance at that frequency.}:
\begin{equation}
  I_{t}(t) = I_{0} + \sum_{i=1}^{3} I_{i} \cos(2i\pi ft + \psi_{n})
\end{equation}
When $\overline{I_{t}} = v_{avg}$, the electrical dynamics of the circuit model 
closely approximate the physical dynamics of a BRS under conditions of \textit{Dynamic Equilibrium}.
$\overline{I_{t}}$ under steady state conditions is proportional to $\overline{V_{t}}$ by Ohms Law.
So that $I_{0} = v_{avg}$ under steady-state conditions, the DC (average) component of Voltage must be:
\begin{equation}
  \overline{V_{t}} = \overline{F_{t}} = mg\sin(\theta) + \frac{1}{2} \rho CdAv_{avg}^2
\end{equation}
\\
The relationship between $\overline{V_{t}}$ and the left, right, and net voltage functions
can thus be derived from Equations \ref{eq:lefttorque}, \ref{eq:righttorque}, and \ref{eq:nettorque}:
\begin{equation}
  V_{l}(t) = \overline{V_{t}}(\frac{1}{2} + \cos(2\pi ft + \frac{\pi}{2}) + \frac{\cos(4\pi ft + \pi)}{2} + \frac{\cos(6\pi ft + \frac{\pi}{2})}{4})
\end{equation}
\begin{equation}
  V_{r}(t) = \overline{V_{t}}(\frac{1}{2} + \cos(2\pi ft - \frac{\pi}{2}+\alpha) + \frac{\cos(4\pi ft + \pi)}{2} + \frac{\cos(6\pi ft - \frac{\pi}{2})}{4})
\end{equation}
\begin{equation}
  V_{t}(t) = \overline{V_{t}}(1 + 2\cos(2\pi ft + \frac{\alpha}{2})\sin(\frac{\alpha}{2}) + \cos(4\pi ft + \pi))
\end{equation}
We are interested in the interaction of the first 
harmonics of Voltage and Current. For the sake of convenience,
phasor notation will be adopted to represent these signals. Subscript $h1$
denotes the electrical quantity as it relates to the first harmonic in Root Mean Square (RMS):
\begin{equation}
  V_{l,h1} = \frac{\overline{V_{t}}}{\sqrt{2}} \angle \frac{\pi}{2}
\end{equation}
\begin{equation}
  V_{r,h1} = \frac{\overline{V_{t}}}{\sqrt{2}} \angle - \frac{\pi}{2} + \alpha
\end{equation}
\begin{equation}
  V_{t,h1} = \sqrt{2}\overline{V_{t}} \sin(\frac{\alpha}{2}) \angle \frac{\alpha}{2}
\end{equation}
The resultant current, from Ohms Law, is:
\begin{equation}
  I_{t,h1} = \frac{V_{t,h1}}{Z_{h1}} = \frac{V_{t,h1}}{X_{h1} + R} = \frac{\overline{V_{t}}\sin(\frac{\alpha}{2})}{\sqrt{2}\sqrt{(\omega m)^2 + R^2}} \angle \frac{\alpha}{2} + \tan^{-1}(\frac{\omega m}{R})
\end{equation}
The amount of work performed can be computed via the dot product of the voltage and current
phasors:
\begin{equation}
  W_{l,h1} = |V_{l,h1}||I_{t,h1}|\cos(\frac{\alpha}{2} + \tan^{-1}(\frac{\omega m}{R}) - \frac{\pi}{2})
\end{equation}
\begin{equation}
  W_{r,h1} = |V_{r,h1}||I_{t,h1}|\cos(\frac{-\alpha}{2} + \tan^{-1}(\frac{\omega m}{R}) + \frac{\pi}{2})
\end{equation}
Noting that $|V_{l,h1}| = |V_{r,h1}| = \frac{\overline{V_{t}}}{\sqrt{2}}$, and $\cos(a) - \cos(b) = -2\sin(\frac{a+b}{2})\sin(\frac{a-b}{2})$
\begin{equation}
  W_{l,h1} - W_{r,h1} = \sqrt{2} \overline{V_{t}}\sin(\tan^{-1}(\frac{\omega m}{R}))\sin(\frac{\alpha-\pi}{2}) |I_{t,h1}|
\end{equation}
Substituting for $|I_{t,h1}|$:
\begin{equation}
  W_{l,h1} - W_{r,h1} = \sqrt{2} \overline{V_{t}} \sin(\frac{\alpha-\pi}{2})\sin(\tan^{-1}(\frac{\omega m}{R}))\frac{\overline{V_{t}}\sin{\frac{\alpha}{2}}}{\sqrt{2}|Z_{h1}|}
\end{equation}
Given that $\sin(\frac{\alpha}{2})\sin(\frac{\pi-\alpha}{2}) = \frac{1}{2}\sin(\alpha)$, and 
substituting for $|Z_{h1}|$:
\begin{equation}
  W_{l,h1} - W_{r,h1} = \frac{\overline{V_{t}}^2}{\sqrt{(\omega m)^2+(\frac{1}{2}\rho CdA v)^2}}\sin(\alpha)\sin(\tan^{-1}(\frac{\omega m}{R}))
\end{equation}
Now:
\begin{equation}
  \epsilon_{balance} = \frac{W_{l,h1}-W_{r,h1}}{\overline{V_{t}I_{t}}} = \frac{mg\sin(\theta) + \frac{1}{2}\rho CdA v^2}{v_{avg}\sqrt{(\omega m)^2+(\frac{1}{2}\rho CdA v)^2}}\sin(\alpha)\sin(\tan^{-1}(\frac{\omega m}{R}))
  \label{eq:error}
\end{equation}
\section{Methods}
This paper will quantitatively assess the impact of system parameters
on power balance error under conditions of dynamic equilibrium
in order to determine if there are scenarios where power balance
is unacceptably high. The system model derived above depends on the following parameters:
\begin{description}
  \item[$v_{avg}$] - The linear velocity of the bicycle-rider system with respect to the ground.
  \item[$\theta$] - The angle of inclination of the cyclist.
  \item[$RPM$] - The \textit{Cadence} of the cyclist, which is related to $\omega$ by $\omega = \frac{RPM * 2 \pi}{60}$.
  \item[$\rho, Cd \& A$] - The air density, the coefficient of drag, and the frontal surface area of the cyclist respectively.
  \item[$m$] - The mass of the entire bicycle-rider system.
  \item[$\alpha$] - As defined in Section \ref{subsec:pedalmodel}.
  \end{description}
Phase Offsets ($\alpha$) of up to $-30^{\circ}$ have been observed 
in competitive cyclists\cite{BINI201156}, however even greater offsets may exist for 
non-competitive or injured cyclists.
The impact of $\alpha$ (varying between $-30^{\circ}$ and $30^{\circ}$ in $1^{\circ}$ increments), and $\epsilon_{balance}$
was assessed under four realistic outdoor riding scenarios ($m = 80kg$), and two low inertia scenarios $m = 4kg$), mimicing an indoor bicycle
ergometer with a flywheel providing inertia. In all scenarios, the kinematic parameters
were set to realistic values - $CdA = 0.3m^{2}$, $g = 9.81ms^{-2}$.
Scenario $6$ represents an indoor riding scenario under high resistance ($\rho$
was adjusted to mimic this effect).\\
\\
Scenario definitions are presented in Table \ref{table:scenarios}.
Simulation was performed using a simple Python Script using
NumPy and Equation \ref{eq:error}. The validity
of the equation derived in \ref{sec:dynamicmodel}
was verified using LtSpice - an time-domain analogue circuit simulation tool
for a small number of cases.
\section{Results}
\begin{figure}
    \centering
  \includegraphics[scale=0.75]{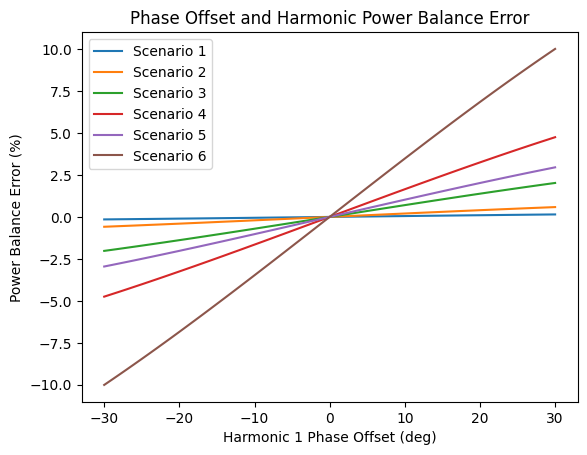}
  \label{fig:simresults}
  \caption{Simulation Results}
  \end{figure}

The worst-case power balance error simulation results are shown in 
Table \ref{table:scenarios} and represented
in their relation to $\alpha$ in Figure \ref{fig:simresults}.\\
Power Balance Error exhibits linearity and symmetry with respect to Phase Offset ($\alpha$).
Power Balance Error is low under low force, high cadence,
and high inertia scenarios, and is greatest under the high
resistance and low inertia indoor bicycle ergometer scenario.
The maximum computed error was more than 10\%, which is substantial
in the context of the stated accuracy of cycling power meters - which
generally claim $\pm 1\%$ accuracy\cite{faveroproduct}\cite{shimanoproduct}\cite{garminproduct}.\\
\begin{table}[h!]
  \caption{Caption text}\label{tab1}%
  \begin{tabular}{@{}l|lllll|ll@{}}
  \toprule
  Scenario & $\rho$ ($kgm^{3}$) & $m$ ($kg$)& $v_{avg}$ ($ms^{-1})$ & $g \sin(\theta)$ $(\%)$ & RPM & $P_{avg}$\footnotemark[1] ($W$) & $\epsilon$\footnotemark[2] (\%)\\
  \midrule
  S1 & 1.204 & 80 & 11 & 0.0 & 80 & 240.379 & 0.148\\
  S2 & 1.204 & 80 & 7 & 0.05 & 70 & 336.626 & 0.586\\
  S3 & 1.204 & 80 & 4 & 0.1 & 60 & 325.478 & 2.023\\
  S4 & 1.204 & 80 & 3 & 0.15 & 50 & 358.036 & 4.749\\
  S5 & 1.204 & 4 & 11 & 0.0 & 80 & 240.379 & 2.954\\
  S6 & 5.0 & 4 & 7 & 0.0 & 60 & 257.25 & 10.008\\
  \botrule
  \end{tabular}
  \footnotetext{A total of 61 simulations were run for each scenario}
  \footnotetext[1]{$P_{avg}$ is computed as $F * v_{avg}$}
  \footnotetext[2]{$\epsilon$ corresponds to the greatest (positive or negative) value of $\epsilon_{balance}$ in the given scenario}
  \label{table:scenarios}  
\end{table}

\pagebreak
\section{Discussion}
The simulation results indicate that substantial power
balance error may occur when a cyclist exhibits a phase
offset. Ostensibly, phase offsets greater that $30^{\circ}$ would result in even greater 
power balance error. There are a number of implications that arise from this finding.\\
\\
Firstly, experiments that aim to measure Power Balance Error, and relate
this measurement to clinical outcomes, must be mindful of the methodology
used to compute this metric, and how this varies from Power Meter to Power Meter.
A second implication of this result is that Power Meters that measure
\textit{Unilateral Power} - that is, power meters that measure torque only on a single
side and provide an estimate of total power by doubling computed power - may produce
erroneous power measurements if phase asymmetry exists. Finally, 
this paper notes that the preponderance of Power Meters
available on the market do not in fact measure \textit{Power Balance}
per se, with the exception of the Favero Assioma - they in fact 
measure a metric akin to \textit{Torque Balance}.
\section{Conclusion}
This paper has analytically derived a relationship
between the phase offset of the primary harmonic component of 
torque in a typical cycling pedal stroke, and Power Balance Error.
This paper demonstrates that realistic deviations in Phase
Symmetry may produce high Power Balance Error under certain
riding scenarios. This paper notes that there may be clinical
implications to this finding.
\section{Statements and Declarations}
The author declares that they have no conflict of interest.
\subsection{Data Availability Statement}
The data used to produce the results in this report 
will be provided upon reasonable request via email.
\bibliography{sn-bibliography}

\end{document}